\documentclass[10pt,conference]{IEEEtran}
\usepackage{amsmath,amssymb,graphicx}
\usepackage{bbm}  
\usepackage{physics} 
\usepackage{algorithm, algorithmic} 
\usepackage{color,xcolor}
\usepackage{subfigure}
\usepackage{psfrag}
\usepackage{cite}
\usepackage{booktabs}
\usepackage{amsmath}
\usepackage{multirow,multicol}
\usepackage{epsfig}
\usepackage{epstopdf}
\usepackage{balance}
\usepackage{mathtools, cuted}
\usepackage{lipsum}
\usepackage{longtable}
\usepackage{array}
\usepackage{booktabs}
\usepackage{tabularx}
\usepackage{booktabs}
\usepackage{enumerate}
\usepackage{graphicx,cite}
\usepackage{times}
\usepackage{amssymb,lipsum}
\usepackage{mathtools}

\begin{document}
\title{Multi-Agent Deep Reinforcement Learning for Optimized Multi-UAV Coverage and Power-Efficient UE Connectivity}
\author{
\IEEEauthorblockN{Xuli Cai, Poonam Lohan, and Burak Kantarci}
\IEEEauthorblockA{\textit{University of Ottawa, Ottawa, ON, Canada}\\
\{xcai049, ppoonam, burak.kantarci\}@uottawa.ca}
\vspace{-0.3in}

}
\maketitle

\begin{abstract}
In critical situations such as natural disasters, network outages, battlefield communication, or large-scale public events, Unmanned Aerial Vehicles (UAVs) offer a promising approach to maximize wireless coverage for affected users in the shortest possible time. In this paper, we propose a novel framework where multiple UAVs are deployed with the objective to maximize the number of served user equipment (UEs) while ensuring a predefined data rate threshold. UEs are initially clustered using a K-means algorithm, and UAVs are optimally positioned based on the UEs' spatial distribution. To optimize power allocation and mitigate inter-cluster interference, we employ the Multi-Agent Deep Deterministic Policy Gradient (MADDPG) algorithm, considering both LOS and NLOS fading. Simulation results demonstrate that our method significantly enhances UEs coverage and outperforms Deep Q-Network (DQN) and equal power distribution methods, improving their UE coverage by up to 2.07 times and 8.84 times, respectively.
\end{abstract}


\section{Introduction}
Wireless communication networks are evolving to support the increasing demand for high data rates and low-latency services \cite{10032494}. Traditional terrestrial infrastructure faces challenges in providing seamless coverage, particularly in remote, disaster-struck, or high-density urban environments. As a result, UAV-assisted communication has emerged as a viable solution to complement existing networks and provide on-demand connectivity. UAVs, functioning as aerial base stations, offer flexibility in deployment, mobility for coverage optimization, and the ability to adapt to dynamic network conditions \cite{parvaresh2023continuous,parvaresh2022tutorial,10177808}. However, effective UAV placement and power allocation remain critical challenges due to interference and fading conditions \cite{10554811}. To address these challenges, we introduce a novel framework for UAV deployment and resource management to maximize the number of UE served at a predefined data rate threshold. Much previous work has applied DRL approaches\cite{10811139,10695917,10412167}, such as DQN \cite{10536060}, DDPG \cite{10628007,9497328}, SAC \cite{10159156}, and PPO \cite{10146333}. We believe that adopting a multi-agent DRL framework could provide a highly effective solution for multi-UAV scenarios.

Our approach begins with the uniform distribution of UE on a grid, followed by  K-means clustering to form UE groups. Each cluster is assigned a UAV, which is optimally positioned based on spatial distribution of UEs. To manage power allocation efficiently and mitigate interference, we employ the MADDPG algorithm, a reinforcement learning (RL)-based technique that enables cooperative decision-making among multiple UAVs. Furthermore, our model incorporates both LOS and NLOS fading effects to ensure realistic channel modeling. The contributions of this work are as follows:
\begin{itemize}
\item Propose a K-means clustering-based approach for UEs grouping, UAVs' allocation, and determine the optimal UAV positions.
\item Integrate MADDPG with dynamic power allocation, improving multi-UAVs coverage efficiency and interference management.
\end{itemize}

Compared to centralized DQN and equal power allocation, the proposed decentralized MADDPG strategy improves UE coverage efficiency by serving up to 2.07 times and 8.84 times more users, respectively.

The rest of the paper is organized as follows: Section II discusses related works. Section III presents the system model and problem formulation. Section IV details the proposed solution methodology, and Section V provides simulation results and performance analysis. Finally, Section VI concludes the paper and outlines future research directions.

\section{Related Work}

The optimization of power allocation and deployment strategies in UAV-assisted systems has garnered significant attention, particularly in mobile edge computing (MEC) scenarios and energy-efficient UAV operations. Several studies have employed RL techniques to address these challenges, focusing on trajectory design, task offloading, and energy management.
\cite{10032494} proposed various MEC frameworks utilizing MADDPG algorithms to optimize task scheduling, trajectory planning, and resource allocation. These studies demonstrated significant improvements in energy efficiency, reduced task processing delays, and fairness in resource distribution across UAVs. Additionally, \cite{9209079} extended these approaches to ensure geographical and load fairness while optimizing energy consumption for UAV-assisted MEC networks.
For energy-efficient UAV path planning, \cite{10433807,10554811} introduced MADDPG-based algorithms that minimized energy usage through techniques like pruning and optimization of neuron layers, as well as by addressing eavesdropping threats in MEC systems with ground-based jamming. 
The study presented in \cite{10584314}  focuses on the joint optimization of content caching probability, resource allocation, and UAV flight trajectory. To achieve this, the authors propose a MADDPG-based Resource allocation and UAV trajectory Optimization (MRFO) algorithm to maximize the overall system energy efficiency. 
In air-ground collaborative networks, \cite{10716801} proposed multi-UAV systems leveraging Lyapunov optimization and MADDPG for adaptive task offloading, service instance management, and resource allocation. These approaches minimized energy consumption and economic expenditure, demonstrating fast convergence and superior cost efficiency compared to baseline methods. Dynamic and adaptive UAV operations were further explored in \cite{10742941,10177808}, where advanced algorithms like MADDPG-LC, Multi-Agent Proximal Policy Optimization (MAPPO), and PPO2-based DRL were employed for dynamic trajectory control, cooperative UAV swarm management, and 3-D trajectory design. These studies highlighted improved energy efficiency, faster convergence, and robustness in addressing flight dynamics and disaster recovery scenarios. \cite{10811139} employs DQN and DDPG to address bandwidth and power allocation for a single UAV operating in a static, interference-free environment. However, cooperation among multiple UAVs is essential for more complex scenarios.

In contrast to previous works that primarily focus on energy efficiency, task offloading, and trajectory planning using RL techniques such as MADDPG, MAPPO, and DDPG, our contributions differentiate our work from the related literature by addressing the challenges of optimal spatial UAV deployment and dynamic interference management by optimally allocating power to UEs in multi-UAV-assisted networks to enhance coverage efficiency by maximizing the number of users served, rather than prioritizing energy efficiency. While the existing literature provides a robust foundation for UAV-assisted wireless communication, several key areas remain for further investigation, such as extending current approaches to address dense urban environments and large-scale UAV deployments and enhanced coverage efficiency by optimizing power usage without compromising network performance.

\section{System Model and Problem Formulation}

\subsection{System Model}
UAVs enhance wireless coverage by providing flexible deployment and connectivity in challenging environments. 
We consider a multi-UAV-assisted communication system, where $N$  UEs are uniformly distributed across a two-dimensional square field $\Psi$ with sides of $L$ meters. The square field consists of $100 \times 100$ grids with each grid cell being a square of side $l=L/100$ meters. In this setup, illustrated in Fig. 1, these users are grouped in different clusters and communicate with a dedicated UAV for each cluster. The number of UAVs deployed are equal to the number of UEs' clusters. Note that the user clustering and UAVs deployment approach is discussed in the next section.  Each UE $UE_i$ is identified by $i\in I{\triangleq}{1,2,\ldots,N}$, and their respective positions are defined by coordinates $(x_m,y_,0)$ relative to the left lower vertex of $\Psi$, $(0,0,0)$. The location of each UAV $UAV_j$ is represented by coordinates $(x_q,y_q,h_q)$ in three-dimensional space, where $h$ represents the hovering height of all UAVs.

Consider a designated UE $i$ depicted in Fig. 1, situated at a horizontal distance $d_{i.j}\triangleq\sqrt{(x_j-x_i)^2+(y_j-y_i)^2}$ from the associated UAV $UAV_j$, and the elevation angle of the UAV $UAV_j$ to that user is $\theta_{i,j}$ rad. For simplicity, we utilize Euclidean distance metrics in our analysis. Given that  UAVs maintain an altitude of $h$ meters above the field $\Psi$, the distance between UE $UE_i$ and the UAV  $UAV_j$ can be calculated as $r_{i,j}\triangleq \sqrt{d_{i,j}^2+h^2}= \frac{h}{sin(\theta_i)}$. All the UAVs and users are assumed to be equipped with a single antenna.

\begin{figure}[t!]
    \centering
\includegraphics[width=8.2 cm]{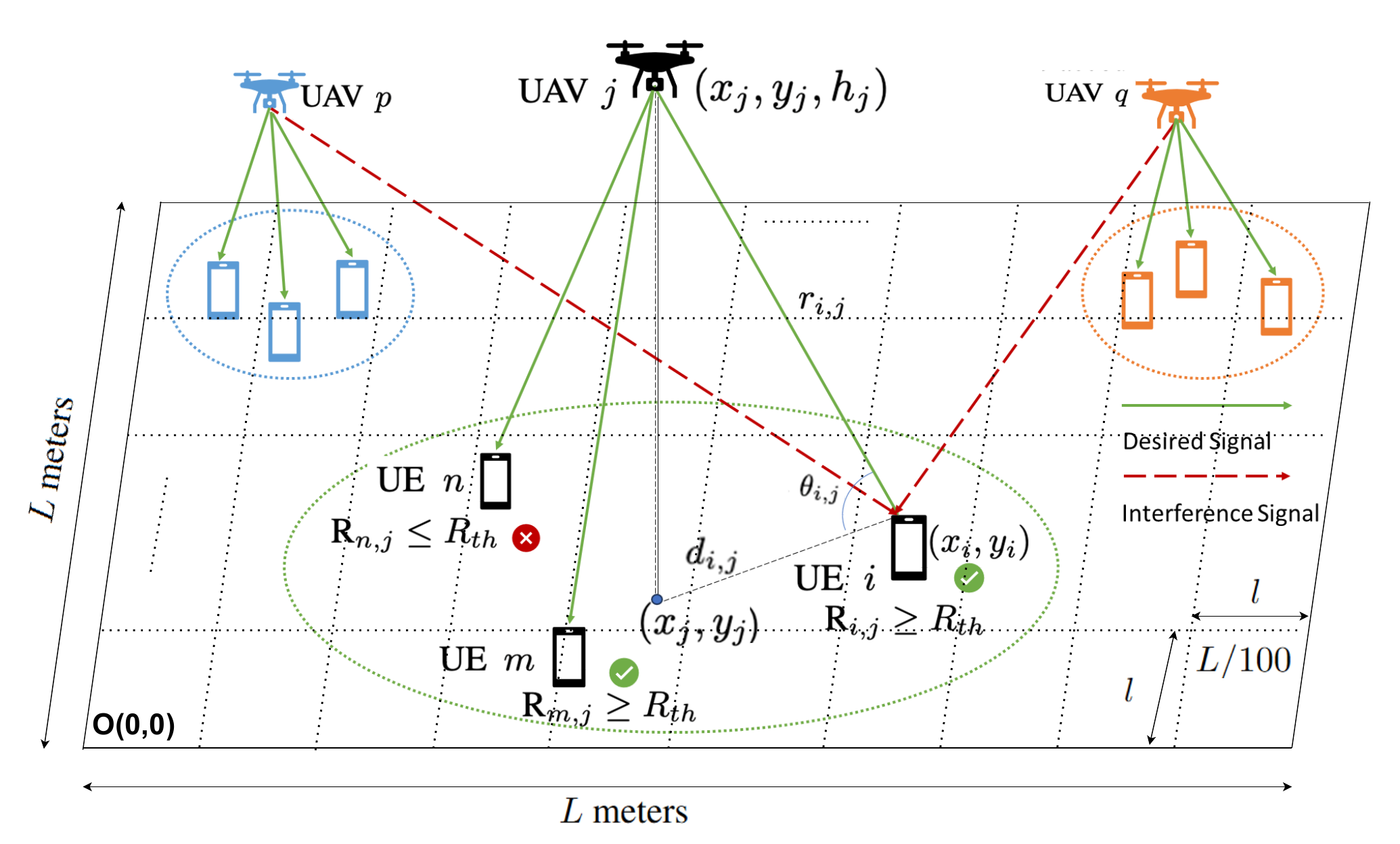}
    \caption{System Model}
    \label{fig:problem}
\end{figure}
 One common approach for air-to-ground channel modeling between the UAV and users is to consider the LoS and NLoS links separately along with their different occurrence probabilities \cite{Hourani}. Note that for NLoS link, the path loss exponent factor $\alpha_{NLoS}$ is higher than that in the LoS link $\alpha_{LoS}$ due to the shadowing effect and reflection from obstacles.  Also, to incorporate the effect of small-scale fading, we are considering Rician fading in LoS links and Rayleigh fading in NLoS links. Consequently, the random channel power gains, $g_i$, for LoS link are noncentral-$\chi^2$ distributed with mean $\mu$ and rice factor $\mathcal{K}$ \cite{simon}, and the random channel power gains, $k_i$, for NLOS link are exponentially distributed with mean $\mu$. Here,  $\mu$ is the average channel power gain parameter that depends on antenna characteristics and average channel attenuation.  With this consideration, the received power for Los and NLos links at  $UE_i$ associated with $UAV_j$ can be written as:
\begin{equation}
   \text{P}^r_{\text{LoS}_{i,j}} = {P_{i,j} g_{i,j} r_{i,j}^{-\alpha_{LoS}}}, \; \forall i\in I,  \quad  
\end{equation}
\begin{equation}
    \text{P}^r_{\text{NLoS}_{i,j}} = { P_{i,j} k_{i,j} r_{i,j}^{-\alpha_{NLoS}}}, \; \forall i\in I.   \
\end{equation}
where $P_{i,j}$ is the transmission power allocated to $UE_i$ by $UAV_j$. The probability of LoS link between  $UE_i$ and $UAV_j$ depends upon the elevation angle $\theta_{i,j}=\sin^{-1}(\frac{h}{r_{i,j}})$, density and height of buildings, and environment. The LoS probability $P_{\text{LoS}_{i,j}}$ is written as \cite{Hourani}:
\begin{equation}\label{eq:plos}
P_{\text{LoS}_{i,j}}={1}/{(1+c\exp(-b[({180}/{\pi})\theta_{i,j}-c]))},
\end{equation} 
where $C$ and $B$ are constants that depend on the environment (rural, urban, dense urban). The probability of NLoS link is $P_{NLoS}=1-P_{LoS}$. Thus the effective SINR received by $UE_i$ associated with $UAV_j$ is expressed as:
\begin{equation}\label{eq:eq1}
    \text{P}^r_{\text{eff}_{i,j}} = P_{\text{LoS}_{i,j}} \cdot \text{P}^r_{\text{LoS}_{i,j}} + P_{\text{NLoS}_{i,j}} \cdot \text{P}^r_{\text{NLoS}_{i,j}}
\end{equation}
Since we are considering the multi-UAV scenario, inter-cluster interference arises from adjacent UAVs transmitting on overlapping frequency bands. Using Shannon's capacity formula, the data rate, $\eta_{i,j}$ bits per sec  (bps) for $UE_i$ through $UAV_j$ communication link  can be expressed as:
 \begin{align}\label{eq:eq2}
R_{i,j}\triangleq (B/N_j) \log_2\left(1+\frac{\text{P}^r_{\text{eff}_{i,j}}}{I_{i,j}+N_o}\right)  \; \forall i\in I.
\end{align}
Here $N_o$ denotes AWGN (additive white Gaussian noise) power, and the inter-cluster interference experienced by $UE_i$ is defined as $I_{i,j}=\sum_{s \neq j}P_{s,avg} k_{i,s} r_{i,s}^{-\alpha_{NLoS}} $. In this expression, only NLoS links are considered for interference calculation, as the probability of a LoS link from interfering UAVs is very low. Since each UAV distributes its entire bandwidth among its associated users, the average transmit power is used as the interfering power. Note that a user $UE_i, \forall i\in I$, is considered under coverage or served by the UAV, if its data rate meets or exceeds the desired rate threshold $R_{th}$, i.e., $R_{i,j} \geq R_{th}$.

\subsection{Problem Formulation}
Following the system model, our objective is to maximize coverage by serving the maximum possible number of users with a given set of multiple UAVs. This objective can be achieved by optimally positioning the multiple UAVs and optimally allocating power resources to their associated users while considering the limited power budget $P_t$ constraints of each UAV and reducing the inter-cluster interference. 
Let $I$ be the set of UEs and $\mathcal{J}$ be the set of UAVs. Define the binary variable in (\ref{eq:binval}).

\begin{equation}
a_{i,j} = 
\begin{cases}
1, & \text{if UE } i \text{ is served by UAV } j, \\
0, & \text{otherwise,}
\end{cases}
\label{eq:binval}
\end{equation}
and let $N_j = \sum_{i\in I} a_{i,j}$ denote the number of UEs served by UAV $j$. The optimization problem is formulated as follows:
\begin{align}
&(\mathcal{P}):\;\max_{(a_{i,j},\, P_j,\, \mathbf{p}_j)} \sum_{j\in\mathcal{J}} \sum_{i\in I} a_{i,j},\;\\
&\text{s.t.:}\; (C1): R_{i,j} \mathbbm{1}(a_{i,j}=1) \ge R_{th},\quad \forall\, i\in I,\, \forall\, j\in\mathcal{J};\nonumber\\
&(C2):\sum_{j\in\mathcal{J}} a_{i,j} \le 1,\quad \forall\, i\in I;\nonumber\\
&(C3):\; a_{i,j} \in \{0,1\},\quad \forall\, i\in I,\, \forall\, j\in\mathcal{J};\nonumber\\
&(C4): \sum_{l=1}^{N_{s,j}} P_{lj} \le P_t; \quad \forall\, j\in\mathcal{J};\nonumber\\
&(C5):\mathbf{p}_j=(x_j,y_j) \in \Psi, \quad \forall\, j\in\mathcal{J}.\nonumber
\end{align}

Constraint (C1) ensures that for UE \(i\) to be considered as a served user, its achieved data rate \(R_{i,j}\)  should be at least the minimum required rate threshold \(R_{th}\). Constraint (C2) guarantees that each UE is served by at most one UAV. Constraint (C3) enforces the binary nature of the variable \(a_{i,j}\), meaning a UE is either served by a UAV or not. Constraint (C4) restricts the total transmit power of each UAV to not exceed its maximum allowable power \(P_t\). Constraint (C5) ensures that the horizontal position of each UAV, given by \(\mathbf{p}_j=(x_j,y_j)\), lies within the feasible region \(\Psi\). To solve this combinatorial and non-convex problem, we present a novel MADDPG-based solution in the next section.

\section{Proposed Methodology}

The proposed methodology comprises two main parts: 1) clustering UEs' locations and determining UAVs' positions using  K-Means and 2) power allocation using MADDPG.

\subsection{Clustering}
We have a set of UE positions, where each UE \(i \in \{1, 2, \ldots, N\}\) is observed once. The UE position for UE \(i\) is denoted as $
\mathbf{u}_i = [x_i,\, y_i] \in \Psi$. Thus, the set of all UEs' positions is given by
$\mathcal{U} = \{ \mathbf{u}_1,\, \mathbf{u}_2,\, \ldots,\, \mathbf{u}_{N} \}
$.

For a given number of clusters \(K \in \{1, 2, \ldots, K_{\max}\}\), the  K-Means clustering problem is formulated as:
\begin{equation}
\min_{\{c_j\}_{j=1}^K,\, \{s(i)\}_{i=1}^{N}} \sum_{i=1}^{N} \left\| \mathbf{u}_i - c_{s(i)} \right\|^2,
\end{equation}
where \(c_j \in \Psi\) denotes the centroid of cluster \(j\), representing a candidate UAV horizontal position and
   \(s(i) \in \{1, 2, \ldots, K\}\) is the cluster assignment for UE \(i\). The solution to the above problem provides the set of centroids \(\{c_1, c_2, \ldots, c_K\}\), which serve as the positions for UAVs and the corresponding UE clusters \(\mathcal{C}_j = \{ \mathbf{u}_i \mid s(i) = j \}\) for \(j = 1, \ldots, K\).


\subsection{MADDPG For Power Allocation}

The UAV power allocation problem is modeled as a multi-agent system, where each UAV is an independent agent interacting with the environment. The objective of each UAV is to maximize the number of served UE within its cluster while minimizing penalties due to data rate oversupply and excessive power usage. The MADDPG algorithm is employed for solving this problem by leveraging a centralized training and decentralized execution framework.

\subsubsection{Reinforcement Learning Problem Formulation}

The problem is defined as a Markov Decision Process (MDP) for \( K \) UAV agents, with the following components:

\paragraph{State Space (\( \mathbf{s} \))}
The state space for each agent \( j \) at timestep \( t \) is defined in (\ref{eq:statespace}) where 
 \( \mathbf{P}_j = \{P_{1j}, P_{2j}, \ldots, P_{N_{s,j}j}\} \) stands for the power allocation to UE within the cluster of UAV \( j \), \( \mathbf{R}_j = \{R_{1j}, R_{2j}, \ldots, R_{N_{s,j}j}\} \) denotes the achieved data rates for UE associated with UAV \( j \), and \( N_{s,j} \) represents the number of UEs in cluster $\mathcal{C}_j$ associated with UAV \( j \).
\begin{equation}
\mathbf{s}_j^t = \{\mathbf{P}_j, \mathbf{R}_j, N_{s,j}\}
\label{eq:statespace}
\end{equation}

\paragraph{Action Space (\( \mathbf{a}_j \))}
The action space for each agent \( j \) corresponds to the adjustment of power allocated to the UE in its cluster as formulated in (\ref{eq:actspace}).
\begin{equation}
\mathbf{a}_j^t = \{\Delta P_{j1}, \Delta P_{j2}, \ldots, \Delta P_{jN_{s,j}}\},
\label{eq:actspace}\end{equation}
where \( \Delta P_{ji} \) represents the change in power allocated to UE \( i \) by UAV \( j \).

\paragraph{Reward Function (\( r) \)}
The reward in (\ref{eq:reward}) is computed as the sum of two components:
the total number of served UEs and the effective total data rate (i.e., the aggregate data rate across all UEs after subtracting the wasted data rate). Let total served users be defined as the sum of the number of served UEs  $(U_{C_j})$ in each cluster $ C_j; \forall j=1,...,k$ (i.e., those all UEs whose effective data rate \(R_{ij}\) meets or exceeds the threshold \(R_{\mathrm{th}}\)) as formulated in the first component of the reward function. The total data rate is formulated in the second summation component in the reward function where $W_d$ denotes the unutilized data rate (i.e., the excess data rate above the required threshold that is not effectively utilized). 
\begin{equation}
r=\sum_{j=1}^{K} U_{C_j} + \left(\sum_{j=1}^{K}\sum_{i=1}^{U_{C_j} } R_{ij} - W_d \right)
\label{eq:reward}
\end{equation}


\paragraph{Transition Dynamics}
The environment transitions from state \( \mathbf{s}_j^t \) to \( \mathbf{s}_j^{t+1} \) based on the UAV’s action \( \mathbf{a}_j^t \). The updated power allocation affects the SINR, data rate, and the resulting reward.

\subsubsection{Centralized Training and Decentralized Execution}

In the MADDPG framework, centralized training is employed using a global critic network, while execution is decentralized using individual actor networks.

\paragraph{Critic Network (\( Q_j \))}
The centralized critic evaluates the joint action-value function in (\ref{eq:critic}) where \( \mathbf{s} \) is the global state, \( \mathbf{a} = \{\mathbf{a}_1, \mathbf{a}_2, \ldots, \mathbf{a}_{K}\} \) is the joint action of all agents, and \( \gamma \) is the discount factor.
\begin{equation}
Q_j(\mathbf{s}, \mathbf{a}) = \mathbb{E}\left[\sum_{t=0}^{T} \gamma^t r_j^t \;|\; \mathbf{s}, \mathbf{a}\right],
\label{eq:critic}
\end{equation}

\paragraph{Actor Network (\( \mu_j \))}
Each agent \( j \) uses an actor network to determine its action as shown in (\ref{eq:action}) where $\theta_{\mu_j}$ are the parameters of the actor network for agent $j$.
\begin{equation}
\mathbf{a}_j^t = \mu_j(\mathbf{s}_j^t | \theta_{\mu_j}),
\label{eq:action}
\end{equation}

\begin{algorithm}[t]
\caption{Proposed MADDPG Solution}
\label{alg:maddpg}
\begin{algorithmic}[1]
    \REQUIRE Number of agents $k$, replay buffer $\mathcal{D}$, batch size $W$, discount factor $\gamma$, target network update rate $\tau$.
    \STATE Initialize actor network $\mu_{\theta_j}$ and critic network $Q_{\phi_j}$ for each agent $j$, with random parameters $\theta_j$ and $\phi_j$.
    \STATE Initialize target networks $\mu_{\theta_j'}$ and $Q_{\phi_j'}$ with weights $\theta_j' \leftarrow \theta_j$, $\phi_j' \leftarrow \phi_j$.
    \STATE Initialize replay buffer $\mathcal{D}$ (shared by all agents).
    \FOR{episode $= 1$ to $M$}
        \STATE Initialize a random process $\mathcal{N}$ for action exploration.
        \STATE Receive initial global state $\mathbf{s}_0$.
        \FOR{$t = 0$ to $T-1$}
            \FOR{each agent $j \in \{1, \ldots, K\}$}
                \STATE Select action $a_j^t = \mu_{\theta_j}(\mathbf{o}_j^t) + \epsilon$, where $\epsilon \sim \mathcal{N}$, and $\mathbf{o}_j^t$ is agent $j$'s local observation.
            \ENDFOR
            \STATE Execute joint action $\mathbf{a}^t = (a_1^t, \ldots, a_K^t)$ in the environment.
            \STATE Check the power limit and prioritize to nearby UE.
            \STATE Collect the power matrix of all UEs, calculate the interference and finalize the data rate matrix.
            \STATE Observe next global state $\mathbf{s}_{t+1}$ and immediate rewards $r_1^t, \ldots, r_K^t$.
            \STATE Store transition $(\mathbf{s}_t, \mathbf{a}^t, r^t, \mathbf{s}_{t+1})$ in $\mathcal{D}$.
            \STATE $\mathbf{s}_t \leftarrow \mathbf{s}_{t+1}$
            \STATE \textbf{If} replay buffer $\mathcal{D}$ has enough samples \textbf{then}
            \STATE \quad \textbf{for} each agent $j$:
            \STATE \quad \quad Sample a mini-batch of $W$ transitions from $\mathcal{D}$.
            \STATE \quad \quad Compute target $y_j = r_j + \gamma\,Q_{\phi_j'}(\mathbf{s}_{t+1}, \mathbf{a}_1^{t+1}, \ldots, \mathbf{a}_K^{t+1})\big\rvert_{a_j^{t+1} = \mu_{\theta_j'}(\mathbf{o}_j^{t+1})}$.
            \STATE \quad \quad Update critic by minimizing loss: 
            \[
               \mathcal{L}(\phi_j) = \frac{1}{W} \sum \bigl( y_j - Q_{\phi_j}(\mathbf{s}_t, \mathbf{a}_1^t, \ldots, \mathbf{a}_K^t)\bigr)^2.
            \]
            \STATE \quad \quad Update actor using policy gradient:
            \[
               \nabla_{\theta_j} J \approx \frac{1}{W}\sum \nabla_{a_j} Q_{\phi_j}(\mathbf{s}_t, \mathbf{a}_t)\big\rvert_{a_j = \mu_{\theta_j}(\mathbf{o}_j^t)} \nabla_{\theta_j} \mu_{\theta_j}(\mathbf{o}_j^t).
            \]
            \STATE \quad \quad Update target networks:
            \[
               \theta_j' \leftarrow \tau \theta_j + (1 - \tau) \theta_j', \quad
               \phi_j' \leftarrow \tau \phi_j + (1 - \tau) \phi_j'.
            \]
        \ENDFOR
    \ENDFOR
\end{algorithmic}
\end{algorithm}

\subsubsection{MADDPG Algorithm}
This proposed MADDPG solution outlined as Algorithm \ref{alg:maddpg}, enables UAVs to learn cooperative policies that maximize the number of served UEs while minimizing penalties for inefficient resource utilization.

\section{Numerical Results}

\subsection{Environment and MADDPG Training Parameters}
Unless otherwise stated all the environmental parameters used in the simulation setup are provided in Table \ref{tab:parameters}. The hyperparameters in MADDPG training are selected to ensure stable and efficient learning. A replay buffer of 100,000 experiences supports off-policy learning, while a batch size of 64 stabilizes updates. A learning rate of 0.0001 is chosen to prevent drastic weight updates and enhance convergence stability, and a discount factor \(\gamma=0.95\) balances short and long-term rewards. A soft update rate \(\tau=0.01\) ensures smooth policy updates, while an exploration noise of \(\sigma_{\text{noise}}=0.2\) promotes exploration, preventing premature convergence to suboptimal policies. The selected values of all hyper-parameters are summarized in Table \ref{tab:maddpg_parameters}.
In our setup, training typically requires over 500 episodes where each episode includes up to 500 time steps, resulting in several hours of GPU computation time. The simulations have been conducted over 10 different seeds, and all figures present the average results across these 10 runs, with $95\%$ confidence interval (CI) bars indicating variability.


\begin{table}[ht]
\centering
\caption{Default Environmental Parameters Used in the Simulations}
\label{tab:parameters}
\begin{tabular}{@{}cll@{}}
\toprule
\textbf{Symbol}         & \textbf{Description}                      & \textbf{Value} \\ \midrule
\(L\)                   & Side of square field $\Psi$                 & \(10000\) meters        \\
\(l\times l\) & Default cell size &\(100 \times 100\) meters\\ 
\(h\) & Height of UAVs & \(500\) meters\\
\(N\)                 & Total Number of UE                    & \(30\)                     \\
\(P_t\)          & Total Power of each UAV                & \(1\) W                    \\
\(B\)                   & Total bandwidth of each UAV                       & \(10\) MHz                  \\
\(R_{\text{th}}\)  & Data Rate Requirement per UE   & \(30\) Mbps                 \\
\(N_o\)            & Noise Power                              & \(4\times10^{-15}\) W             \\
\(\alpha_{\text{LOS}}\) & Path Loss Exponent for LoS               & \(3\)                   \\
\(\alpha_{\text{NLOS}}\)& Path Loss Exponent for NLoS              & \(4\)                   \\
\(c\)                   & Environmental Constant \text{(Dense Urban)} & \(11.95\)                \\
\(b\)                   & Environmental Constant \text{(Dense Urban)}              & \(0.136\)                 \\
\(\mathcal{K}\) & Fading Factor & \(10\)\\
\(\mu\) & Mean Power & \(0.5\) \\
\bottomrule
\end{tabular}
\end{table}

\begin{table}[ht]
\centering
\caption{MADDPG Hyperparameters}
\label{tab:maddpg_parameters}
\begin{tabular}{@{}cll@{}}
\toprule
\textbf{Symbol}         & \textbf{Description}                           & \textbf{Value} \\ \midrule
\(B_s\)                 & Replay Buffer Size                             & \(100,000\) samples   \\
\(W\)                 & Batch Size for Training                        & \(64\) samples        \\
\(\alpha_a\)            & Actor Network Learning Rate                    & \(0.0001\)             \\
\(\alpha_c\)            & Critic Network Learning Rate                   & \(0.0001\)             \\
\(\gamma\)              & Discount Factor for Future Rewards             & \(0.95\)              \\
\(\tau\)                & Target Network Update Rate                     & \(0.01\)              \\
\(\sigma_{\text{noise}}\)& Exploration Noise Level                        & \(0.2\)               \\
\(H\)                   & Hidden Layer Sizes for Actor and Critic        & \([128, 128]\)        \\
\bottomrule
\end{tabular}
\end{table}

\noindent

\begin{table}[htbp]
\caption{Mean Transmission Power Usage}

\centering
\begin{tabular}{cccc}
\hline
\textbf{Clusters} & \textbf{MADDPG} & \textbf{DQN} & \textbf{Equal Power} \\ \hline
5  & 98.81\% & 99.00\% & 100\% \\
10 & 40.53\% & 46.00\% & 100\% \\
15 & 7.19\%  & 10.27\% & 100\%  \\
20 & 1.73\%  & 3.85\%  & 100\%  \\
25 & 0.89\%  & 2.80\%  & 100\%  \\ \hline
\end{tabular}
\label{tab:mean_cluster_power}
\end{table}

\begin{figure}[ht] \centering
    \includegraphics[width=8cm]{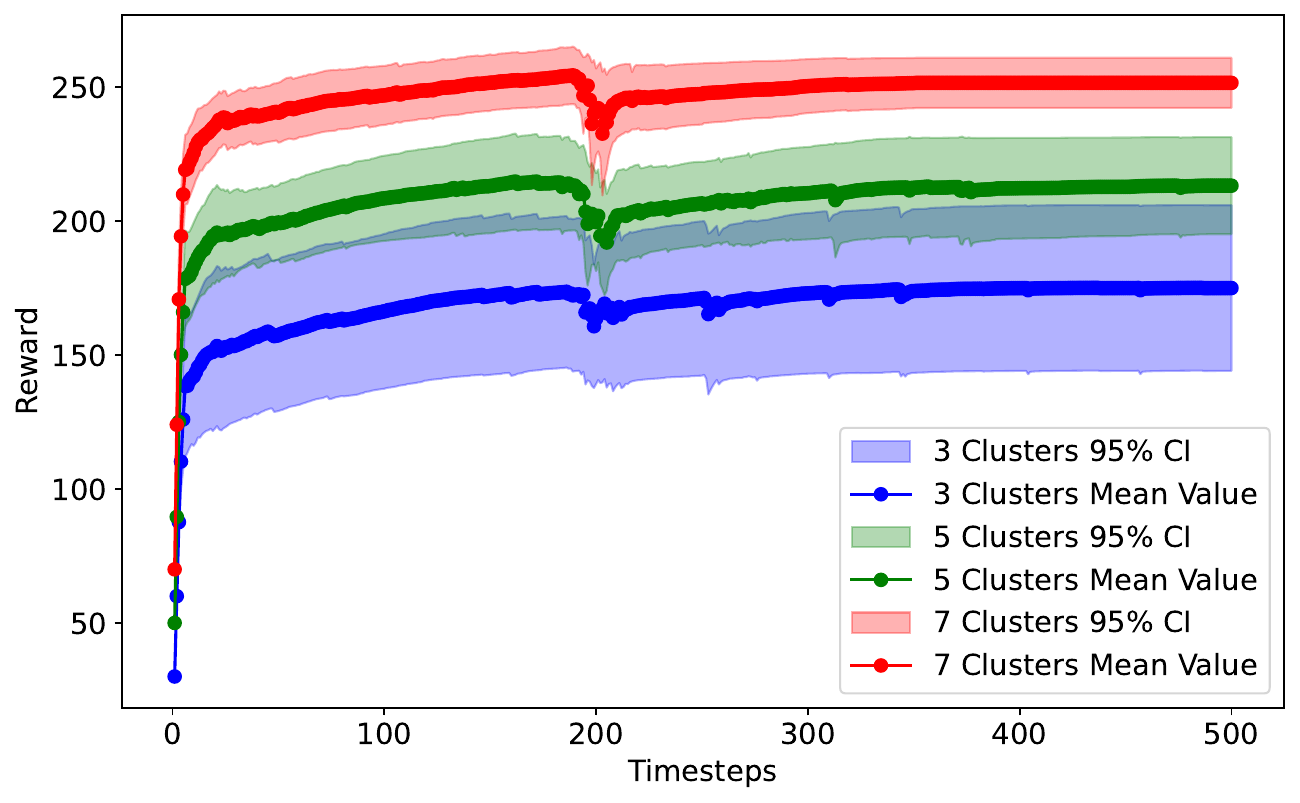} 
    \caption{MADDPG Training Convergence ($N=30$ UEs, $R_{th}=30$ Mbps)}
    \label{fig:training_plot}
\end{figure}



\begin{figure*}[ht]
    \centering
    \includegraphics[width=\textwidth]{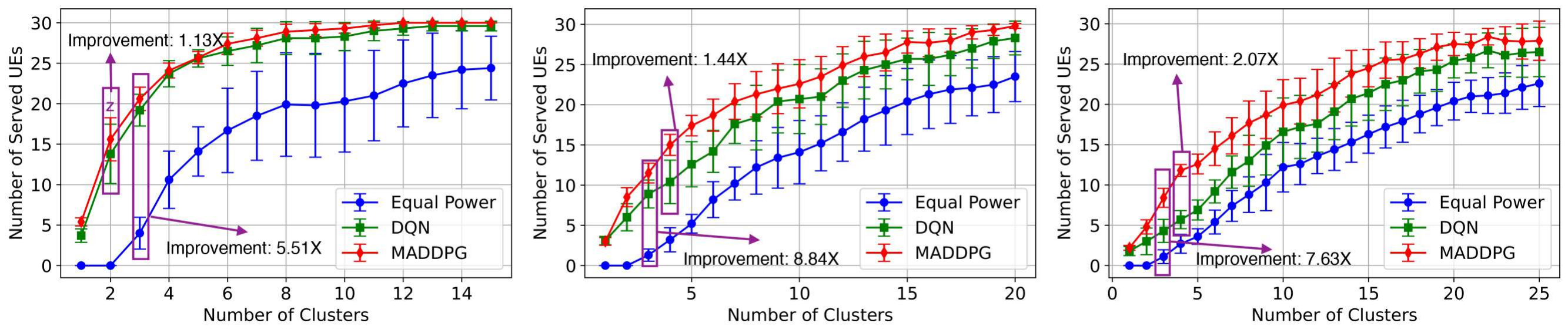}\vspace{-0.3cm}
    \caption{Performance comparison for $N=30$ UEs: (a) $R_{th}=10$ Mbps (b) $R_{th}=20$ Mbps (c) $R_{th}=30$ Mbps}
    \label{fig:combined_vertical_plots}
\end{figure*}

\begin{figure}[ht] \centering
    \includegraphics[width=8cm]{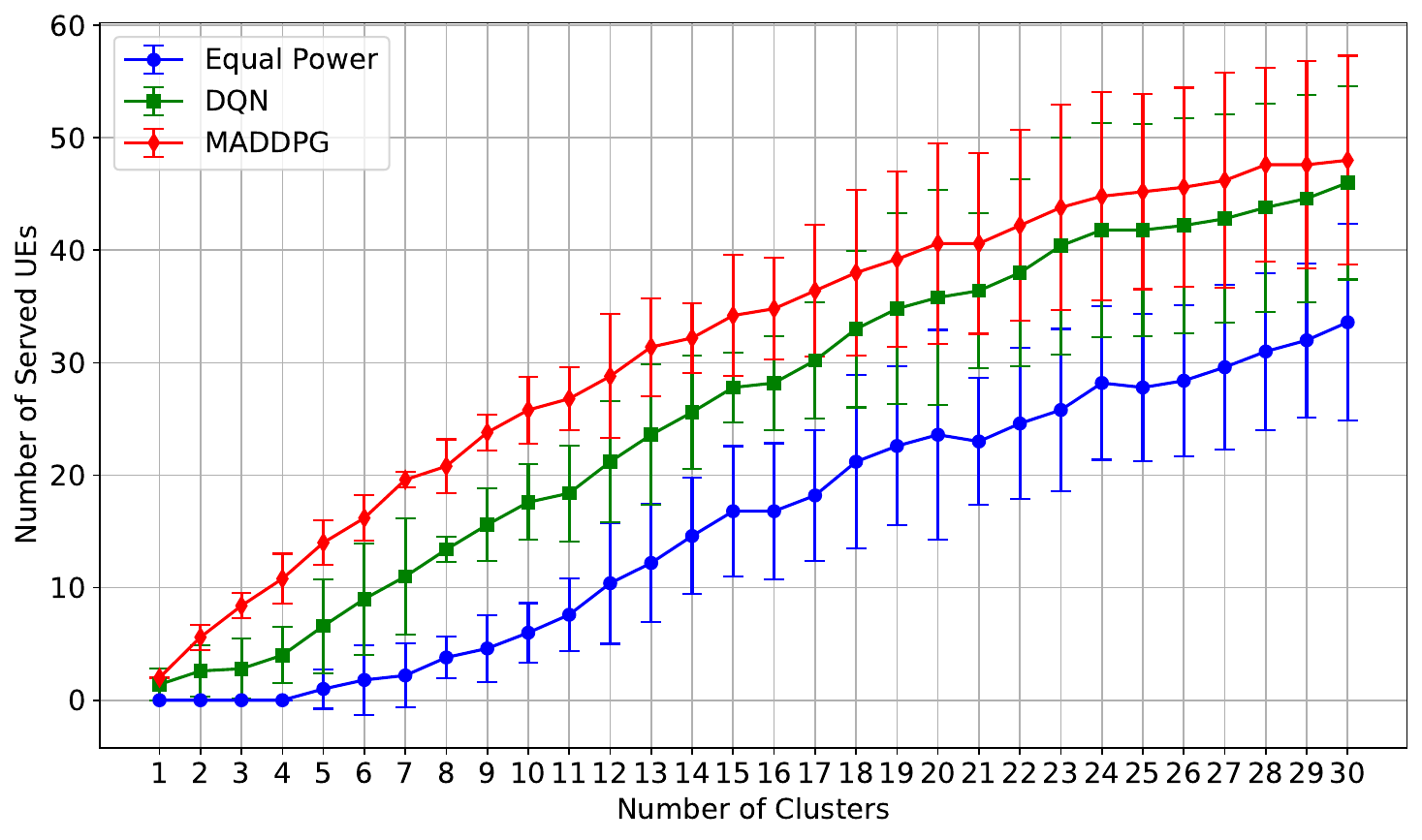} 
    \caption{Performance comaprison for $N=60$ UEs ($R_{th}=30$ Mbps)}
    \label{fig:served_plot_60_vehicles}
\end{figure}

\begin{figure}[ht] \centering
    \includegraphics[width=8cm]{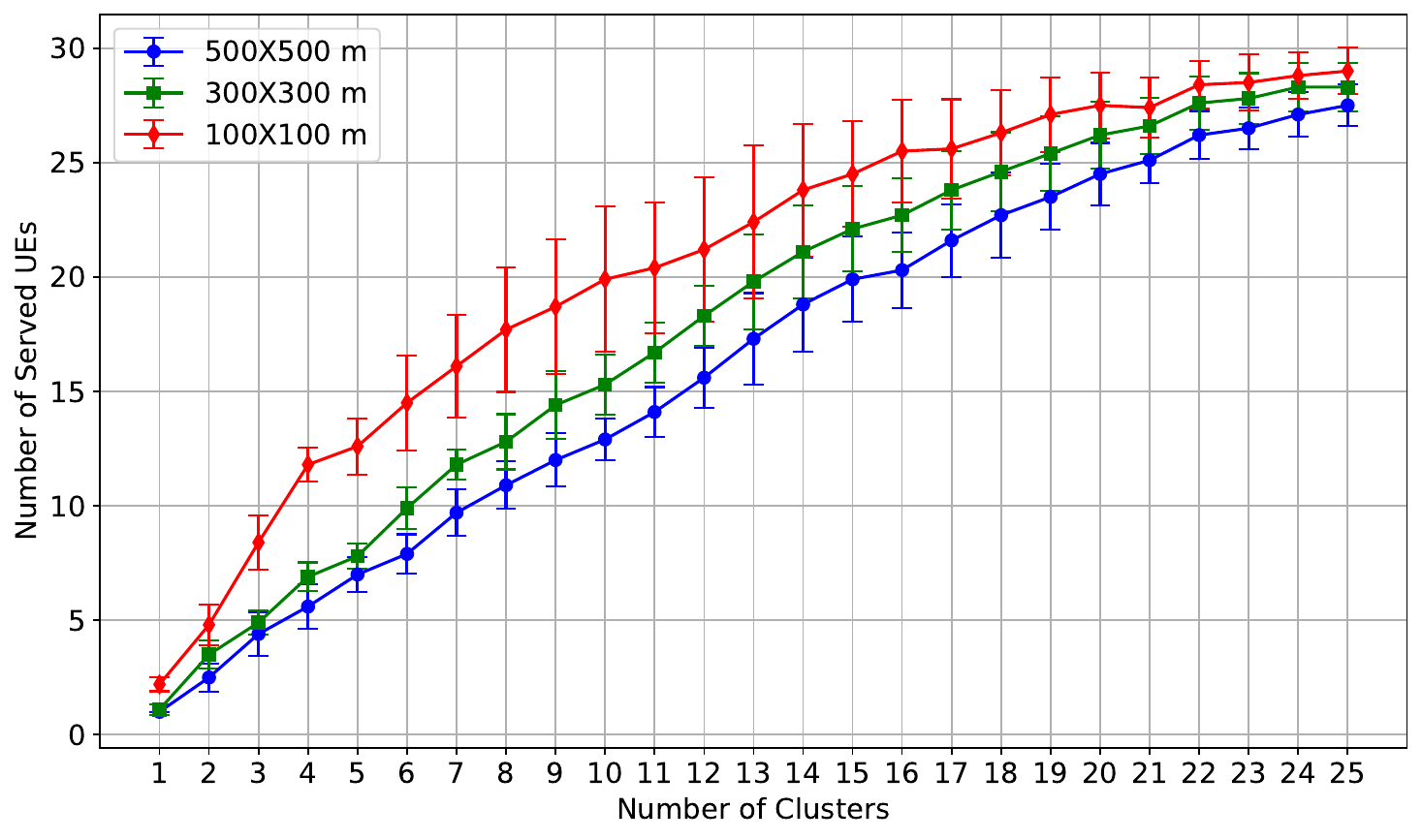} 
    \caption{MADDPG performance for different cell scales (N=30, $R_{th}$=30 Mbps)}
    \label{fig:served_plot_30_block_scale_diff}
\end{figure}

\subsection{Results and Discussion}
\subsubsection{Training Convergence}
Fig.~\ref{fig:training_plot} shows the convergence behavior of MADDPG training over $500$ time-steps for $3$, $5$, and $7$ clusters. The reward starts low and steadily increases, stabilizing after $100$ time-steps. Minor drops likely result from policy updates or exploration-exploitation trade-offs. Training with 7 clusters yields the most stable policy with minimal variance and the highest reward, suggesting that more clusters enhance learning through richer interactions and exploration.


\subsubsection{Served UEs Comparison}
Fig.~\ref{fig:combined_vertical_plots} compare the performance of the proposed MADDPG solution against DQN and equal power allocation. To ensure a fair comparison, three different $R_{th}$ values are considered. The results indicate that as $R_{th}$ increases, the performance gap between MADDPG and DQN widens, particularly for a smaller number of clusters, leading to a higher number of served users with the proposed approach. Also, with less $R_{th}$  requirements, fewer clusters/UAVs are sufficient to cover all the users. Furthermore, Fig.~\ref{fig:served_plot_60_vehicles} shows the comparison results with higher user density (with N=60 UEs) indicating better performance of MADDPG over DQN and equal power allocation.

\subsubsection{Power Usage Comparison}
Table~\ref{tab:mean_cluster_power} presents the average transmission power consumption of UAVs for MADDPG, DQN, and equal power allocation across varying cluster counts. With fewer clusters, UAVs must cover larger distances to serve users, resulting in higher transmission power requirements. As the number of clusters increases, power consumption decreases since users are distributed more evenly, reducing the distance between UAVs and their associated users. Among the evaluated approaches, MADDPG demonstrates better performance with lower transmission power usage.

\subsubsection{Cell Scale Impact}
Fig.~\ref{fig:served_plot_30_block_scale_diff} shows MADDPG performance across different cell scales/coverage areas with L=10000 m, L=30000 m, and L=50000 m. Larger coverage areas/cell scales result in greater distances between UEs and their associated UAV clusters, leading to a lower number of served users. However, as the number of clusters/UAVs increases, this performance gap narrows since UAVs are positioned closer to UEs, improving coverage efficiency.

\section{Conclusion}
This paper has presented a multi-UAV-assisted wireless network proposing K-means clustering and MADDPG-based solution for optimal positioning of UAVs and optimal power allocation, respectively. Compared to centralized DQN and equal power distribution, our decentralized MADDPG approach improves UE coverage efficiency maximum of 2.07 times and 8.84 times, respectively. The framework incorporates realistic LoS/NLoS fading and interference modeling, accurately capturing wireless dynamics. By leveraging MADDPG, UAVs autonomously learn optimal strategies and enhance UE coverage, as well as data rates while maximizing network performance. Our ongoing research aims to further enhance the system's adaptability by incorporating dynamic UE mobility models and tackling energy-efficient trajectory optimization . 


\section*{Acknowledgements}
This work is supported in part by National Science and Engineering Research Council (NSERC) Discovery and NSERC CREATE TRAVERSAL programs.
\bibliographystyle{IEEEtran}

\end{document}